\documentstyle[12pt]{article}
\catcode`\@=11 \@addtoreset{equation}{section}

\catcode`\@=12

\begin{document}

\title{The Decay Lifetime of Polarized Fermions in Flight}
\author{Zhi-qiang Shi
\thanks{Present address: Residence 10-2-7, Shaanxi Normal University, Xi'an 710062,
China. E-mail address: zqshi@snnu.edu.cn}\\
{\small\it{Faculty of Science, Xi'an Jiaotong University, Xi'an 710049, China}}
\\ Guang-jiong Ni
\thanks{E-mail address: gj\_ni@Yahoo.com}\\
\small\it{Department of Physics, Fudan University, Shanghai, 200433, P.R.China}\\
\small\it{Department of Physics, Portland State University, Portland, OR 97207 USA.}}
\date{}

\maketitle
\begin{abstract}
Based on the parity violation in Standard model, we study the dependence of lifetime on
the helicity of an initial-state fermion in weak interactions. It is pointed out that if
the initial fermions in the decays are longitudinally polarized, then the decay lifetime
 of left-handed polarized fermions is different from that of right-handed
polarized fermions in flight with a same velocity in a same inertial system.
\\[0.4cm] PACS numbers: 11.30.Er; 11.30.Rd;12.15.-y; 13.30.-a; 13.35.Bv
\end{abstract}

\hskip\parindent

The phenomena of parity violation are experimentally exhibited in two aspects. (1) The
angular distribution of the particles emitted from decay processes is asymmetric. The
experiment has established that the emission of beta particles is more favored in the
direction opposite to that of the $^{60}$Co nuclei spin.[1] In muon decays, this
asymmetry has also been observed.[2] The parity violation in the $\Lambda$-decay is
manifested as an up-down asymmetry of the decay pion (or proton) relative to the
production plane. It is worthy of special mention that the asymmetry of the angular
distribution is dependent on the spin orientation of the initial decay particle. If the
average spin component of the initial particle is equal to zero, the angular distribution
is symmetric. (2) All of fermions emitted from decay processes are longitudinally
polarized. It is well known that the neutrinos are left-handed (LH) polarized while the
antineutrinos are right-handed (RH) polarized. In beta or muon decays it is found that
the electrons are always LH polarized while the positrons are RH polarized.[3] Based on
the two experimental facts above, naturally, it is thought that not only the final-state
fermions in the decays, but also the initial-state fermion should be relevant to its
longitudinal polarization. So we further could consider that the lifetime of fermions in
the LH helicity state and the that of fermions in the RH helicity state should be
different as well. However, The experimental and theoretical study on the polarization or
helicity of the initial fermions in decays has not yet been discussed fully in the
literature. In this paper we aim at exploring this point in some detail.

\section{The charged weak currents in the SM}
\hskip\parindent In order to describe the parity violation in weak interactions, all of
fundamental fermions are divided into two classes, LH chirality state and RH chirality
state, in the standard model (SM). They are defined as
$$\psi_{_L}=\frac{1}{2}(1+\gamma_{_5})\psi,\quad
\psi_{_R}=\frac{1}{2}(1-\gamma_{_5})\psi,\eqno{(1)} $$
respectively. And we have
$$\psi=\psi_{_L}+\psi_{_R}.\eqno{(2)}$$
The LH chirality state is different from the RH chirality state. The former is the
SU(2)-doublet state whereas the latter the SU(2)-singlet state, hence they have different
gauge transformations. Especially, RH chirality state has zero weak isospin and is only
present in neutral weak currents. Therefore, there exist the only LH chirality states in
charged weak currents. The interaction Lagrangians for charged weak lepton current and
charged weak quark current read, respectively
$$
{\cal L}_{\ell\:W}= \frac{1}{\sqrt{2}}\:g_{_2}\:{\overline e_{_L}}\gamma_\mu
W_\mu^+\nu_{_L} +\frac{1}{\sqrt{2}}\:g_{_2}\:{\overline \nu_{_L}}\gamma_\mu
W_\mu^-e_{_L}, \eqno{(3)}
$$
$$ {\cal L}_{q W}= \frac{1}{\sqrt{2}}\:g_{_2}\:{\overline d_{_L}}\gamma_\mu W_\mu^+u_{_L}
+\frac{1}{\sqrt{2}}\:g_{_2}\:{\overline u_{_L}}\gamma_\mu W_\mu^-d_{_L},\eqno{(4)}
$$
where $g_{_2}$ is the coupling constant corresponding to SU(2). Obviously, all of
fermions are in the LH chirality states and all of antifermions while in the RH chirality
states in decay processes.

For example, the weak interaction in muon decay is successfully described by four-fermion
interaction Hamiltonian. We denote the matrix element by
$$M\sim\sum g^\gamma_{_{\varepsilon\mu}}\langle\overline e_\varepsilon|\Gamma_\gamma|
(\nu_e)_n\rangle\langle(\overline\nu_\mu)_m|\Gamma_\gamma|\mu_\mu\rangle,\eqno{(5)}$$
where $\gamma=S,V,T$ indicates a scalar, vector or tensor interaction; and $\varepsilon,
\mu=R,L$ indicate a right- or left-handed chirality of the electron or muon. The
chiralities $n$ and $m$ of the $\nu_e$ and $\overline\nu_\mu$ are then determined by the
values of $\gamma, \varepsilon$ and $\mu$. All the coupling constants have been obtained
entirely from experiments without any model assumption.[4] The experiments on muon decay
show $g_{_{RL}}$, $g_{_{RR}}$, $g_{_{LR}}$ to be zero, and at least one of the two
coupling, $g^V_{_{LL}}$ or $g^S_{_{LL}}$, to be nonzero. The experiments on inverse muon
decay provides a lower limit for pure $V-A$ interaction with $|g^V_{_{LL}}|>0.960$. Thus
the measurements give a strong support to the standard model which sets $g^V_{_{LL}}=1$
and all the others being zero, and then indicate that the charged weak current is
dominated by a coupling to left-handed chirality fermions. Therefore, the negative muon
decay can be written as
$$\mu_{_L}^-\longrightarrow e_{_L}^{-}+\overline\nu_{e{_R}}+
\nu_{\mu{_L}}. \eqno{(6)}$$
And the matrix element (5) has the form
$$ M=\frac{g_{_2}}{4m^2_W}(\overline\nu_{\mu_
L}\gamma_\mu\:\mu_{_L})(\overline
e_{_L}\gamma_\mu\:\nu_{e_L}).\eqno{(7)}$$

Similarly, the neutron decay can be written as
$$ n_{_L}\longrightarrow p_{_L}+ e_{_L}+ \overline \nu_{_R}, \eqno{(8)}$$
and the corresponding matrix element is
$$
M=\frac{g_{_2}}{4m^2_W}(\overline p_{_L}\gamma_\mu\:n_{_L})(\overline
e_{_L}\gamma_\mu\:\nu_{e_L}).\eqno{(9)}$$

\section{Helicity and chirality}
\hskip\parindent The Dirac equation has the form in Pauli metric
$$(\gamma_\mu\partial_\mu+mc^2)\psi=0,\eqno{(10)}$$
where $m$ is the rest mass and $\psi$ is four-component spinor. The projection of spin
vector $\vec{\sigma}$ along the direction of fermion momentum is known as the helicity or
polarization:
$$h=\frac{\vec{\sigma}\cdot\vec p}{\mid\vec{p}\mid},\eqno{(11)}$$
where $h$ is a constant of the motion with eigenvalues $\pm 1$. Taking the simplest case
of $\vec p$ : $p_z=p$, when $h=+1$ and $h=-1$, we have the RH helicity state
$\psi_{_{Rh}}$ and the LH helicity state $\psi_{_{Lh}}$, respectively
$$\psi_{_{Rh}}=\left(\begin{array}{c}1\\0\\
\frac{p\:c}{E+mc^2}\\0\end{array}\right)e^{^{\frac{i}{\hbar}(p\:z-Et)}},\quad
\psi_{_{Lh}}=\left(\begin{array}{c}0\\1\\0\\
\frac{-p\:c}{E+mc^2}\end{array}\right)e^{^{\frac{i}{\hbar}(p\:z-Et)}}
\eqno{(12)}$$

Setting
$$\psi=\left(\begin{array}{c}\varphi\\\chi\end{array}\right),\eqno{(13)}$$where
$\varphi$ and $\chi$ are two-component spinors, we may rewrite (10) as a set of two
coupled equations for $\varphi$ and $\chi$:
$$\bigg\{\begin{array}{c}\vec\sigma\cdot\vec
p\:c\chi=(E-mc^2)\varphi,\\ \vec\sigma\cdot\vec
p\:c\varphi=(E+mc^2)\chi.\end{array}\eqno{(14)}$$ We have then
$$\varphi=\frac{\vec\sigma\cdot\vec
p\:c}{E-mc^2}\chi,\quad\chi=\frac{\vec\sigma\cdot\vec
p\:c}{E+mc^2}\varphi.\eqno{(15)}$$

On the other hand, $\psi_{_L}$ and $\psi_{_R}$ are the eigenfunctions of the chirality
operator $\gamma_{_5}$, satisfying respectively
$$\gamma_{_5}\psi_{_L}=\psi_{_L},\quad
\gamma_{_5}\psi_{_R}=-\psi_{_R}.\eqno{(16)}$$For describing the relation between the
chirality state and the helicity state, we introduce two linear combination function of
$\varphi$ and $\chi$
$$\phi_{_R}=\frac{1}{\sqrt 2}\:(\varphi+\chi),\quad
\phi_{_L}=\frac{1}{\sqrt 2}\:(\varphi-\chi).\eqno{(17)}$$ Then
$$\varphi=\frac{1}{\sqrt 2}\:(\phi_{_R}+\phi_{_L}),\quad \chi=\frac{1}{\sqrt
2}\:(\phi_{_R}-\phi_{_L}).\eqno{(18)}$$
Substituting into formula (13), we get
$$\psi=\left(\begin{array}{c}\varphi\\\chi\end{array}\right)=\frac{1}{\sqrt
2}\left(\begin{array}{c}\phi_{_L}\\-\phi_{_L}\end{array}\right)+\frac{1}{\sqrt
2}\left(\begin{array}{c}\phi_{_R}\\\phi_{_R}\end{array}\right).\eqno{(19)}$$
Comparing with Eq.(2) this gives
$$\psi_{_L}=\frac{1}{\sqrt
2}\left(\begin{array}{c}\phi_{_L}\\-\phi_{_L}\end{array}\right),\quad
\psi_{_R}=\frac{1}{\sqrt 2}\left(\begin{array}{c}\phi_{_R}\\
\phi_{_R}\end{array}\right).\eqno{(20)}$$
Hence, instead of Eq.(14), we find the Dirac
equation in Weyl (chiral) representation:
$$\bigg\{\begin{array}{c}(E-\vec\sigma\cdot\vec p\:c)\phi_{_R}=mc^2\phi_{_L},\\
(E+\vec\sigma\cdot\vec p\:c)\phi_{_L}=mc^2\phi_{_R}.\end{array}\eqno{(21)}$$ Obviously,
both $\phi_{_L}$ and $\phi_{_R}$ are not the eigenfunction of helicity operator $h$. When
$m=0$ or $E\gg m$, however, they are the eigenfunction of helicity operator $h$. For
example, because neutrino has zero mass, the helicity state of neutrino is identical with
its chirality state.

In general, for $h=-1$, we obtain from Eq.(21)
$$\left|\frac{\phi_{_R}}{\phi_{_L}}\right|_{Lh}
=\frac{mc^2}{E+p\:c}=\sqrt{\frac{c-v}{c+v}}=W,\eqno{(22)}$$ where $v$ is the fermion
velocity and $W$ is the Weyl parameter introduced in [5]. Once $v>0$, a LH polarized
fermion has $h=-1$ explicitly and $\phi_{_L}>\phi_{_R}$ implicitly. One can see that the
hidden amplitude of RH chirality state decreases with the increases of $v$ until
$\phi_{_R}\rightarrow 0$ when $v\rightarrow c$, showing that a high-energy fermion can be
LH polarized without hidden RH spinning instability. With $v=0$, we have
$\phi_{_R}=\phi_{_L}$, i.e. the hidden amplitude of LH chirality state is equal to that
of RH chirality state and one even can not discriminate its polarization being LH or RH
[6].

Similarly, for $h=+1$, we obtain
$$\left|\frac{\phi_{_R}}{\phi_{_L}}\right|_{Rh}
=\frac{mc^2}{E-p\:c}=\sqrt{\frac{c+v}{c-v}}=\frac{1}{W}.\eqno{(23)}$$ Once $v>0$, a RH
polarized fermion has $h=1$ explicitly and $\phi_{_R}>\phi_{_L}$ implicitly. The latter
decreases with the increase of $v$ until $\phi_{_L}\rightarrow 0$ when $v \rightarrow c$,
showing that a high-energy fermion can be RH polarized without hidden LH spinning
instability.

\section{The lifetime of polarized fermions}
\hskip\parindent Usually, in calculating the decay probability $\Gamma$ of a fermion
(muon or neutron) in weak interactions, one defined the lifetime $\tau=1/\Gamma$ with
$\Gamma$ being that of fermion at rest and its spin direction being averaged. So no
discrimination between helicities was involved. Now we will take the velocity dependence
of $\Gamma$ for polarized fermion into account. To this purpose, the absolute square of
spinor $\psi$ reads
$$\mid \psi
\mid^2=\mid\phi_{_L}\mid^2+\mid\phi_{_R}\mid^2=\mid\phi_{_L}\mid^2
\left(1+\frac{\mid\phi_{_R}\mid^2}{\mid\phi_{_L}\mid^2}\right).\eqno{(24)}$$ Which leads
to $$\mid\phi_{_L}\mid^2=\frac{\mid\psi\mid^2}
{1+\left|\frac{\phi_{_R}}{\phi_{_L}}\right|^2}.\eqno{(25)}$$ Since the charged weak
current originates from the left-handed chirality state only, from Eqs. (3) and (4) or
the matrix element (7) and (9), we get the ratio of $\mid\phi_{_L}\mid^2$ in a RH
helicity state to that in a LH helicity state, yielding the relevant ratio of decay
probability for moving fermion in weak interactions,
$$\frac{\Gamma_R}{\Gamma_L}=\frac{\mid\phi_{_L}\mid^2_{_{Rh}}}
{\mid\phi_{_L}\mid^2_{_{Lh}}}=\frac{1+W^2}
{1+\frac{1}{W^2}}=W^2=\frac{c-v}{c+v}.\eqno{(26)}$$ From
$$\Gamma=\frac{1}{2}\:(\:\Gamma_R+\Gamma_L)\quad \hbox{and} \quad
\Gamma=\frac{\hbar}{\tau},\eqno{(27)}$$ we find the RH polarized fermion lifetime
$\tau_{_R}$ and LH polarized fermion lifetime $\tau_{_L}$, respectively
$$\tau_{_R}=\tau\frac{c}{c-v}\quad \hbox{and}\quad
\tau_{_L}=\tau\frac{c}{c+v},\eqno{(28)}$$where $\tau$ is the average lifetime. According
to the theory of special relativity, we may write $\tau$ as
$$\tau=\frac{\tau_{_0}}{\sqrt{1-(v/c)^2}},\eqno{(29)}$$ where $\tau_{_0}$ is the
lifetime of fermion at rest. Therefore
$$\tau_{_R}=\frac{\tau_{_0}}{\sqrt{1-(v/c)^2}}\:\frac{c}{c-v}\quad \hbox{and}\quad
\tau_{_L}=\frac{\tau_{_0}}{\sqrt{1-(v/c)^2}}\:\frac{c}{c+v}.\eqno{(30)}$$ It is easy to
see that the $\tau_{_R}$ is greater than $\tau_{_L}$, which shows the lifetime asymmetry
of left-right handed polarized fermion. The lifetime asymmetry is simply expressed by the
familiar parameter $\beta$ in special relativity
$$\hbox{lifetime asymmetry}\equiv\frac{\tau_{_R}-\tau_{_L}}{\tau_{_R}+\tau_{_L}}=\beta=\frac{v}{c}.\eqno{(31)}$$
When $v=0$, we find $$\tau_{_R}=\tau_{_L}=\tau_{_0},\quad
(\beta=0).\eqno{(32)}$$ When $v\rightarrow c$, we find
$$\tau_{_R}\rightarrow\infty, \quad
\tau_{_R}\rightarrow\infty,\quad (\beta\rightarrow 1).\eqno{(33)}$$ In particular, when
$v=\frac{1}{2}\:c, \:\beta=\frac{1}{2}$, the lifetime of the LH polarized fermions has a
minimum value $\tau_{_L}=\tau_{min}=0.77\tau_{_0}$.

It is not difficult to prove that the lifetimes of antifermions in motion are given by
$$\overline\tau_{_R}=\frac{\tau_{_0}}{\sqrt{1-(v/c)^2}}\:\frac{c}{c+v}\quad
\hbox{and}\quad \overline
\tau_{_L}=\frac{\tau_{_0}}{\sqrt{1-(v/c)^2}}\:\frac{c}{c-v}.\eqno{(34)}$$

\section{Summary and discussion}
\hskip\parindent The Dirac equation is invariant under a space reflection $\vec x
\rightarrow -\vec x$, provided the spinor field $\psi(\vec x,t)$ undergoes the
corresponding transformation
$$\psi(\vec x,t)\rightarrow \psi(-\vec x,t)=\gamma_{_4} \psi(\vec x,t).\eqno{(35)}$$
Applying (35) to (1) we get
$$\bigg\{\begin{array}{c}
\psi_{_L}(\vec x,t)=\frac{1}{2}(1+\gamma_{_5})\psi(\vec x,t)\rightarrow
\frac{1}{2}(1+\gamma_{_5})\psi(-\vec x,t)=\frac{1}{2}(1+\gamma_{_5})\gamma_{_4}
\psi(\vec x,t)=\gamma_{_4}\psi_{_R}(\vec x,t),\\[0.2cm]
\psi_{_R}(\vec x,t)=\frac{1}{2}(1-\gamma_{_5})\psi(\vec x,t) \rightarrow
\frac{1}{2}(1-\gamma_{_5})\psi(-\vec x,t)=\frac{1}{2}(1+\gamma_{_5})\gamma_{_4} \psi(\vec
x,t)=\gamma_{_4}\psi_{_L}(\vec x,t).\end{array}\eqno{(36)}$$ It means that despite the
invariance of Dirac equation under transformation (35), the LH and RH chirality
components of $\psi$ transform each other. Alternatively, it would be better to consider
the space inversion transform here in terms of two-component spinors $\phi_{_L}$ and
$\phi_{_R}$ regardless of the theory being invariant or not after the following
transformation:
$$\bigg\{\begin{array}{c}
\phi_{_L}(\vec x,t)\rightarrow\phi_{_L}(-\vec x,t)\rightarrow\phi_{_R}(\vec x,t),\\
[0.2cm]\phi_{_R}(\vec x,t)\rightarrow\phi_{_R}(-\vec x,t)\rightarrow\phi_{_L}(\vec x,t).
\end{array}\eqno{(37)}$$
It is easy to see that under the space inversion the Dirac equations (21) is invariant,
showing that the law of motion of a Dirac particle preserves the parity symmetry. On the
other hand, the equations (28) and (34) are also invariant, respectively, under a space
reflection and the transformation (37). It means that the $\tau_{_L}$ is always smaller
than $\tau_{_R}$ for fermions in flight and the $\tau_{_R}$ is always smaller than
$\tau_{_L}$ for antifermions in any one of inertial systems. It is just because only the
LH chirality components $(\phi_{_L})$ of fermions participates in the decay process
whereas their RH chirality components $(\phi_{_R})$ do not. For example, under the space
inversion (37), a moving fermion with $h=-1$ having $\phi_{_L}>\phi_{_R}$ and
$\tau_{_L}=\tau/(1+\beta)$ will transform into a fermion moving in opposite direction
with $h=+1$ having $\phi_{_R}>\phi_{_L}$ and $\tau_{_R}=\tau/(1-\beta)$. Hence the
property that the lifetime $\tau_{_L}$ of the LH polarized fermion is shorter than the
lifetime $\tau_{_R}$ of the RH polarized fermion in flight ( with a same speed in the
same inertial system ) shows a serious violation of parity symmetry. More intuitively,
since a neutron having a spin $\vec S_n$ will emit an electron $e^-$ and an antineutrino
$\overline\nu_e$ basically in the backward direction to $\vec S_n$, we may imagine a
decaying neutron being like a "comet" with its head oriented to $\vec S_n$ and its tail
composed of $e^-$ and an $\overline\nu_e$. Obviously, this neutron is by no means an
isotropic object in $\beta$ decay. Rather, it is just like a highly anisotropic "comet".
A space inversion would transform it into an unexisting (ridiculous) "comet" with its
tail parallel to $\vec S_n$, showing the parity violation. Moreover, if we push the rest
"comet" into motion either along with or opposite to the direction of $\vec S_n$, we will
find left-right asymmetry of the lifetime in its decay, showing $\tau_{_L}<\tau_{_R}$
with the same speed. To our understanding, it is just this strange phenomenon relevant to
the parity violation that was overlooked in the past.

As seen from formula (31), the lifetime asymmetry $\beta$ increases with $v$ increases,
and is negligible at low energy. In neutron decay, for example, $\beta=0.046$ when
neutron kinetic energy $T=1$ MeV, and $\beta=0.145$ when $T=10$ MeV. Published neutron
lifetime were measured using thermal or cold neutron beams.[7] The energy of these beams
is below 1 eV, and too small for the lifetime asymmetry effect to be observe. One
actually lacks experimental evidence in support of this rather natural consequence of the
SM or pure V-A theory. In view of the important implications of these observations, we
report them here now in the hope that they may stimulate and encourage further
experimental investigations on the question of polarized fermion lifetime in either beta
or muon decays.

Since $\tau_R>\tau_L$, an unpolarized neutron beams would gradually turn into RH
polarized in propagation process. Due to neutron decay, according to formula (8), the
amount of LH protons and LH electrons would be predominant in the final state. In fact,
as early as 1956 T.D.Lee and C.N.Yang pointed out:``There must exist two kinds of protons
$p_R$ and $p_L$, the right-handed one and the left-handed one. Furthermore, at the
present time the protons in laboratory must be predominantly of one kind in order to
produce the supposedly observed asymmetry".[8] If the lifetime asymmetry of polarized
fermion in flight described above is correct, one kind of helicity fermions would be more
abundant than another kind in some region of the universe. The measured polarized fermion
lifetime may be crucial for further understanding of the weak interactions and there
might be important consequences for cosmology and astrophysics. Especially, it would be
closely related to various neutrino anomalies.

\end{document}